\documentclass[showpacs,preprintnumbers,amsmath,amssymb,onecolumn,nofootinbib]{revtex4}

\usepackage{graphicx}
\usepackage{dcolumn}
\usepackage{bm}

\newcommand{\pr}{Phys. Rev. }

\newcommand{\Apj}{Astrophys. J. }

\begin{document}

\title{Reply to 'Comment on 'Heavy element production in inhomogeneous big bang nucleosynthesis''}
\author{Shunji Matsuura}
\affiliation{Department of Physics, School of Science, University of Tokyo,
7-3-1 Hongo, Bunkyo, Tokyo 113-0033, Japan}

\author{Shin-ichirou Fujimoto }
\affiliation{Department of Electronic Control, 
Kumamoto National College of Technology, Kumamoto 861-1102, Japan}

\author{Masa-aki Hashimoto}
\affiliation{Department of Physics, School of Sciences, 
Kyushu University, Fukuoka 810-8560, Japan}

\author{Katsuhiko Sato}
\affiliation{Department of Physics, School of Science, University of Tokyo,
7-3-1 Hongo, Bunkyo, Tokyo 113-0033, Japan}
\affiliation{Research Center for the Early Universe, University of Tokyo, 
7-3-1 Hongo, Bunkyo, Tokyo 113-0033, Japan}

\begin{abstract}

This is a reply report to \cite{Rauscher:2006mv}.
We studied heavy element production in the high baryon density region in the early universe\cite{Matsuura:2005rb}.
However it is claimed by \cite{Rauscher:2006mv} that a small scale but high baryon density region contradicts observations for the light element abundance or, in order not to contradict to the observations the high density region must be so small
that it cannot affect the present heavy element abundance.

In this paper we study big bang nucleosynthesis in the high baryon density region
and show that in certain parameter spaces it is possible to produce enough amount of the heavy element without contradiction 
to cosmic microwave background and light element observations.

\end{abstract}

\maketitle

\section{Introduction}


In a standard scenario, big bang nucleosynthesis (BBN) can produce only light elements, up to $^{7}$Li, and 
all heavy elements have been synthesized in stars.
However, many phase transitions in the early universe could have printed their trace in a non-standard way. 
For example, some baryogenesis models\cite{Dolgov:1992pu} predict very high baryon density islands in ordinary low density 
backgrounds.

In the previous paper\cite{Matsuura:2005rb}, we studied heavy element production in inhomogeneous BBN from this point of view.
\footnote{For previous works on the inhomogeneous 
 big bang nucleosynthesis, see~\cite{IBBN1, IBBN2}. Heavy elements production is also mentioned in~\cite{jedam}.}
However we limited ourselves to the heavy element abundance and did not discuss about the light element abundance and consistency with observations.
This is because we assumed that high baryon density region is very local and 
do not affect the global light element abundance.
In \cite{Rauscher:2006mv}, Rauscher pointed out that in order not to contradict to observations, the high baryon density region
must be very small and cannot affect the present heavy element abundance.
In this
paper, we show that there is a parameter region in which the heavy element can be produced enough to affect observation while keeping the light element abundance
consistent with observations.
We consider that the disagreement between Rauscher's opinion and our opinion comes from two points.
One is that we are looking at some parameter regions in which neutrons in high baryon density
do not diffuse so much as to cause  disaster  in standard BBN. 
We would like to emphasize this point.
The other is that the relevant quantity is not the spatial size of the high baryon density region but the amount of baryon in high density regions.

We will discuss the following issues:
In section \ref{LEA}, we discuss the light element abundance in the homogeneous high baryon density region and after mixing the high and the low baryon density region.
In section \ref{HEAVY}, we study the heavy element(Ru,Mo) abundance in high and averaged baryon density
and show that heavy elements can be produced enough without contradicting the light element observation.
In section \ref{DIFF}, we briefly comment on the diffusion scale of the high baryon density region.

\section{Light element abundance}
\label{LEA}

\subsection{Homogeneous BBN}
We calculate homogeneous BBN with various values of $\eta$(baryon photon ratio).
In Table.\ref{light-table} and \ref{light-normal}, we show the numerical result of the mass fraction and the number fraction of each light element for $\eta=10^{-3}$ and $3.162 \times 10^{-10}$.
\begin{center}
\begin{table}[htbp]
\begin{tabular}{ccc}\hline
\multicolumn{3}{c}{$\eta=10^{-3}$} \\ \hline
name              & mass fraction                  & number fraction \\
H                 &    $  5.814 \times 10^{-1} $   & $ 8.475 \times 10^{-1}$ \\
$^{4}$He          &  $ 4.185 \times 10^{-1}$       & $ 1.525 \times 10^{-1}$ \\
$^{3}$He          &   $  4.842 \times 10^{-13} $   & $ 1.614 \times 10^{-13}$ \\
$^{7}$Li+$^{7}$Be & $  1.559\times 10^{-12}$       &  $ 2.227 \times 10^{-13}$ \\ 
D                 & $ 1.577 \times 10^{-22}$       & $  7.883\times 10^{-23}$ \\ \hline
\end{tabular}
\caption{The mass and the number fractions of light elements for the homogeneous BBN with $\eta=10^{-3}$}
\label{light-table}
\end{table}
\end{center}
\vspace{0.7cm}

\begin{center}
\begin{table}[htbp]
\begin{tabular}{ccc}\hline
\multicolumn{3}{c}{$\eta=3.162 \times 10^{-10}$} \\ \hline
name & mass fraction & number fraction \\
H                    &   $ 7.58 \times 10^{-1}$       & $ 9.26 \times 10^{-1}$   \\
$^{4}$He             &   $ 2.419 \times 10^{-1}$       & $ 7.39 \times 10^{-2}$  \\
$^{3}$He             &  $ 4.299 \times 10^{-5}$        &$ 1.433 \times 10^{-5}$   \\
$^{7}$Li + $^{7}$Be  &  $ 8.239 \times 10^{-10}$        & $ 1.177 \times 10^{-10}$  \\ 
D                    & $ 1.345 \times 10^{-4}$         &$ 6.723 \times 10^{-5}$   \\ \hline
\end{tabular}
\caption{The mass and the number fractions of light elements for the homogeneous BBN with $\eta=3.162 \times 10^{-10}$}
\label{light-normal}
\end{table}
\end{center}
\vspace{0.7cm}

As baryon density becomes higher, more protons and neutrons are bounded to form $^{4}$He.
At $\eta=10^{-3}$, most of the final product of $^{7}\rm{Li}$ comes from $^{7}\rm{Be}$ which decays to $^{7}\rm{Li}$ after BBN.
Details on light element production for various $\eta$ can also be found in \cite{Wagoner:1966pv}. In this paper we almost concentrate on a case in which the high baryon density 
region has $\eta=10^{-3}$.
We expect that compared to $\eta \geq 10^{-3}$, the profile of the abundance for $\eta=10^{-3}$ is more different from standard BBN because most of the light element abundances change monotonically with respect to $\eta$ and if this 
case does not contradict to observations, other cases would also be consistent.
Briefly, the amount of $\rm{H}$ decreases and $^{4}\rm{He}$ increases monotonically 
as $\eta$ become larger.
The number fraction of $\rm{D}$ is less than $10^{-20}$ for $\eta$ greater than $10^{-7}$.
For $^{3}\rm{He}$, the number fraction drastically decreases around $\eta =10^{-4}$ down to
${\cal O}(10^{-13})$, and for $^{7}\rm{Li}$, the number fraction increases until $\eta=10
^{-6}$ and drastically decreases for a larger value of $\eta$. In the following sections, we will see that this non-standard setup does not strongly 
contradict to the observations. 
For simplicity we ignore the diffusion effect before and during BBN, and
after BBN both high and low baryon density regions are completely mixed.
Detailed analysis such as the case in which the high baryon density region doesn't completely mixed, or
taking into account diffusion effects are left for future work.

\subsection{Parameters and Basic equations}
In this section, we summarize the relations among parameters.

Notations : 
$n$, $n ^{H}$, $n ^{L}$ are averaged, high, and low baryon number density.
$f ^{H}$, $f ^{L}$ are the volume fractions of the high and the low baryon density region.
$y_{i} $, $y_{i} ^{H}$, $y_{i} ^{L}$ are the mass fractions of each element (i) in averaged-, high- and low-density regions.
The basic relations are
\begin{eqnarray}
f ^{H}+f ^{L}&=&1  \label{vol} \\
f ^{H}n ^{H}+f ^{L}n ^{L}&=&n \label{density} \\
y_{i} ^{H}f ^{H}n ^{H}+y_{i} ^{L}f ^{L}n ^{L}&=&y_{i}n. \label{spec}
\end{eqnarray}
Under the assumption that the temperature of the universe is homogeneous, the above equation can be written as
\begin{eqnarray}
f ^{H}\eta ^{H}+f ^{L}\eta ^{L}&=&\eta\label{denseta} \\ 
y_{i} ^{H}f ^{H}\eta ^{H}+y_{i} ^{L}f ^{L}\eta ^{L}&=&y_{i}\eta\label{speceta}
\end{eqnarray}
where $\eta =\frac{n}{n_{\gamma}}$,$\eta ^{H,L} =\frac{n ^{H,L}}{n_{\gamma}}$
Conventional parameters for inhomogeneous BBN are $\eta$, $f$ and density ratio $R=\frac{n^{H} }{n^{L} }$.
Here we use a different combination of parameters.
Relevant values for the abundance analysis are products $f^{H,L} \times \eta ^{H,L} $ and
$\eta ^{H,L}$.
$f^{H,L}_{v}\times\eta ^{H,L}$ determines the amount of baryon from high- and low- density regions.
$\eta ^{H,L}$ determines the mass fraction of each species of nuclei. 
For convenience, we write the ratio of baryon number contribution from high density region 
as $a$, i.e., $f ^{H}\eta ^{H} : f ^{L}\eta ^{L}=a:(1-a) $.
There are 5 parameters($n^{H,L} , n$ and $f^{H,L} $) and 2 constraints
(Eq.(\ref{vol}) and Eq.(\ref{density})).
We calculate the light element abundance for various values of $\eta ^{H,L}$.
$\eta$ can also take any value, but in order not
to contradict observational constraints, we choose $\eta$ from $3.162 \times 10^{-10}$ to $10^{-9}$.
$a$ is determined by Eq(\ref{denseta}). 
The aim of the analysis in this section is not to find parameter regions which precisely agree  with the observational light element abundance and $\eta$ from CMB. 
Our model is too simple to determine the constraints to parameters.
For example, we completely ignore the diffusion effect before and during BBN.
Instead we see that at least our analysis in previous paper is physically reasonable.

\subsection{Theoretical predictions and observations of light elements}

We consider the cases of
$\eta ^{H}=10^{-3}$ and $\eta ^{L}=3.162\times 10^{-10}$.
The mass fractions of and $\rm{H}$ and $^{3}\rm{He}$ in the high density region 
are $0.5814$ and $4.842\times 10^{-13}$, respectively, 
while those in the low density region are $0.758$ and $4.299\times 10^{-5}$.
From Eq.(\ref{speceta}), we have 
\begin{eqnarray}
f ^{H}\eta ^{H}y_{^3\rm{He}} ^{H}+f ^{L}\eta ^{L}y_{^3\rm{He}} ^{L}&=&\eta y_{^3\rm{He}} \\
4.842\times 10^{-13}\times a+4.299\times 10^{-5}\times(1-a)&=& y_{^3\rm{He}}
\end{eqnarray}
\begin{eqnarray}
f ^{H}\eta ^{H}y_{\rm{H}} ^{H}+f ^{L}\eta ^{L}y_{\rm{H}} ^{L}&=&\eta y_{\rm{H}} \\
0.5814\times a+0.758\times(1-a)&=& y_{\rm{H}}.
\end{eqnarray}

\begin{figure}[htbp]
\includegraphics[width=9cm,clip]{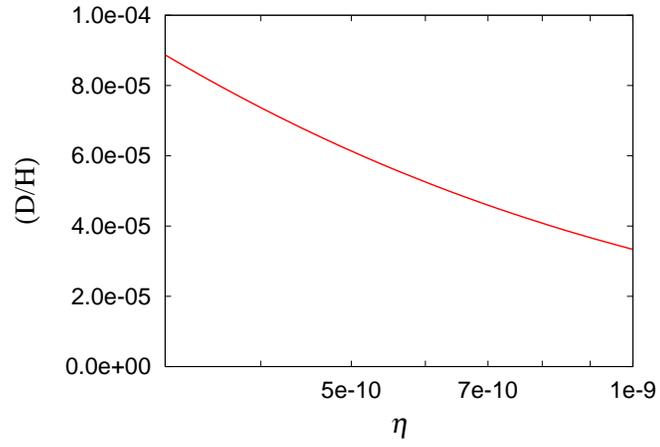}
\caption{Averaged ratio of D to H,(D/H) vs $\eta$}
\label{deps}
\end{figure}

\begin{figure}[htbp]
\includegraphics[width=9cm,clip]{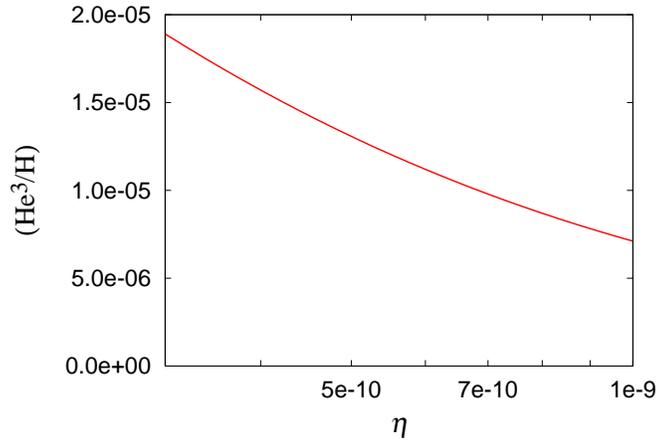}
\caption{Same as Fig.\ref{deps} but for ($^3$He/H)}
\label{heeps}
\end{figure}

\begin{figure}[htbp]
\includegraphics[width=9cm,clip]{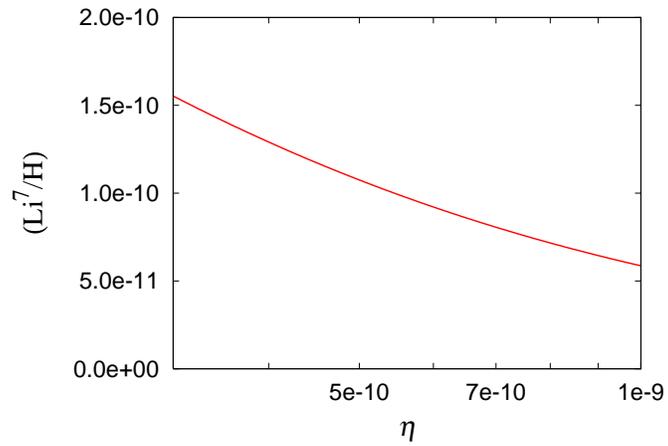}
\caption{Same as Fig.\ref{deps} but for ($^{7}$Li/H)}
\label{lieps}
\end{figure}

We can calculate an averaged value of the abundance ratio of $^3$He to H as

\begin{equation}
(\frac{^3\rm{He}}{\rm{H}})=\frac{1}{3}\frac{4.842\times 10^{-13}\times a+4.299\times 10^{-5}\times(1-a)}
{0.5814\times a+0.758\times(1-a)}.
\end{equation}
where $a$ is related to $\eta$ as 
\begin{eqnarray}
a&=&\frac{\eta ^{H}}{\eta}\frac{\eta - \eta ^{L}}{\eta ^{H}-\eta ^{L}}\\
&=& \frac{10^{-3}}{\eta}\frac{\eta - 3.162\times10^{-10}}{10^{-3}-3.162\times10^{-10}}\\
&\sim&\frac{\eta - 3.162\times10^{-10}}{\eta}.
\end{eqnarray}
Here $a$ varies from 0 to 0.9 for reasonable values of $\eta$, or 
$3.162 \times 10^{-10}-10^{-9}$.
Similarly, for $\eta ^{H}=10^{-3}$
the number fractions are
\begin{equation}
(\frac{\rm{D}}{\rm{H}})=\frac{1}{2}\frac{1.577\times 10^{-22}\times a+1.345\times 10^{-4}\times (1-a)}{0.5814\times a+0.758\times(1-a)}
\end{equation}

\begin{equation}
(\frac{^{7}\rm{Li}}{\rm{H}})=\frac{1}{7}\frac{1.559\times 10^{-12}\times a +8.239\times 10^{-10}\times (1-a)}
{0.5814\times a+0.758\times(1-a)}.
\end{equation}
Fig.\ref{deps},\ref{heeps} and \ref{lieps} represent the averaged abundance ratio,
(D/H), ($^{3}$He/H) and ($^{7}$Li/H) respectively.

We can see that the light element abundance is the same order around 
$\eta \sim 5\times 10^{-10}-10^{-9}$ as observations 
\cite{Fields:2004cb,Kirkman:2003uv,O'Meara:2000dh,Kirkman:1999zu,Linsky:2003ia,Ryan:1999jq,Bonifacio:2002yx,Pinsonneault:2001ub}.
\begin{eqnarray}
(\frac{\rm{D}}{\rm{H}})_{obs}=(1.5-6.7) \times 10^{-5} \label{obs1}\\
(\frac{^{7}\rm{Li}}{\rm{H}})_{obs}=(0.59-4.1)\times 10^{-10}.
\label{obs2}
\end{eqnarray}

We do not discuss detail about diffusion here. But at least above result suggest
that our analysis is not beside the point.



\section{Theoretical predictions and observations of heavy elements ($^{92,94}$\rm{Mo}, $^{96,98}$\rm{Ru})}
\label{HEAVY}

The same analysis can be applied for heavy elements such as $^{92}\rm{Mo}$, $^{94}\rm{Mo}$, $^{96}\rm{Ru}$ and $^{98}\rm{Ru}$.
We are interested in these elements because in many models of
supernovae nucleosynthesis, these p-nuclei are less produced.
We will see that some amount of these heavy elements can be synthesized in BBN.

\begin{center}
\begin{table}[htbp]
\begin{tabular}{cc}\hline
\multicolumn{2}{c}{$\eta=10^{-3}$} \\ \hline
name & mass fraction  \\
H          &    $ 5.814 \times 10^{-1} $  \\
$^{4}$He   &       $4.185\times10^{-1}$ \\
$^{92}$Mo & $1.835\times 10^{-5}$  \\
$^{94}$Mo & $4.1145\times 10^{-6}$  \\
$^{96}$Ru & $1.0789 \times 10^{-5}$  \\
$^{98}$Ru & $1.0362\times 10^{-5}$  \\\hline
\label{heavy}
\end{tabular}
\caption{The mass fractions of nuclei for homogeneous BBN with $\eta=10^{-3}$}
\end{table}
\end{center}

\vspace{0.7cm}

From Table.\ref{heavy}, we can derive the expected value of these elements.

\begin{equation}
(\frac{^{92}\rm{Mo}}{\rm{H}})=\frac{1}{92}\frac{1.835\times 10^{-5}\times a}{0.5814\times a+0.758\times(1-a)}
\end{equation}
\begin{equation}
(\frac{^{94}\rm{Mo}}{\rm{H}})=\frac{1}{94}\frac{4.1145\times 10^{-6}\times a}{0.5814\times a+0.758\times(1-a)}
\end{equation}
\begin{equation}
(\frac{^{96}\rm{Ru}}{\rm{H}})=\frac{1}{96}\frac{1.0789 \times 10^{-5}\times a}{0.5814\times a+0.758\times(1-a)}
\end{equation}
\begin{equation}
(\frac{^{98}\rm{Ru}}{\rm{H}})=\frac{1}{98}\frac{1.0362\times 10^{-5}\times a}{0.5814\times a+0.758\times(1-a)}.
\end{equation}

\vspace{1cm}

\begin{figure}[htbp]
\includegraphics[width=9cm,clip]{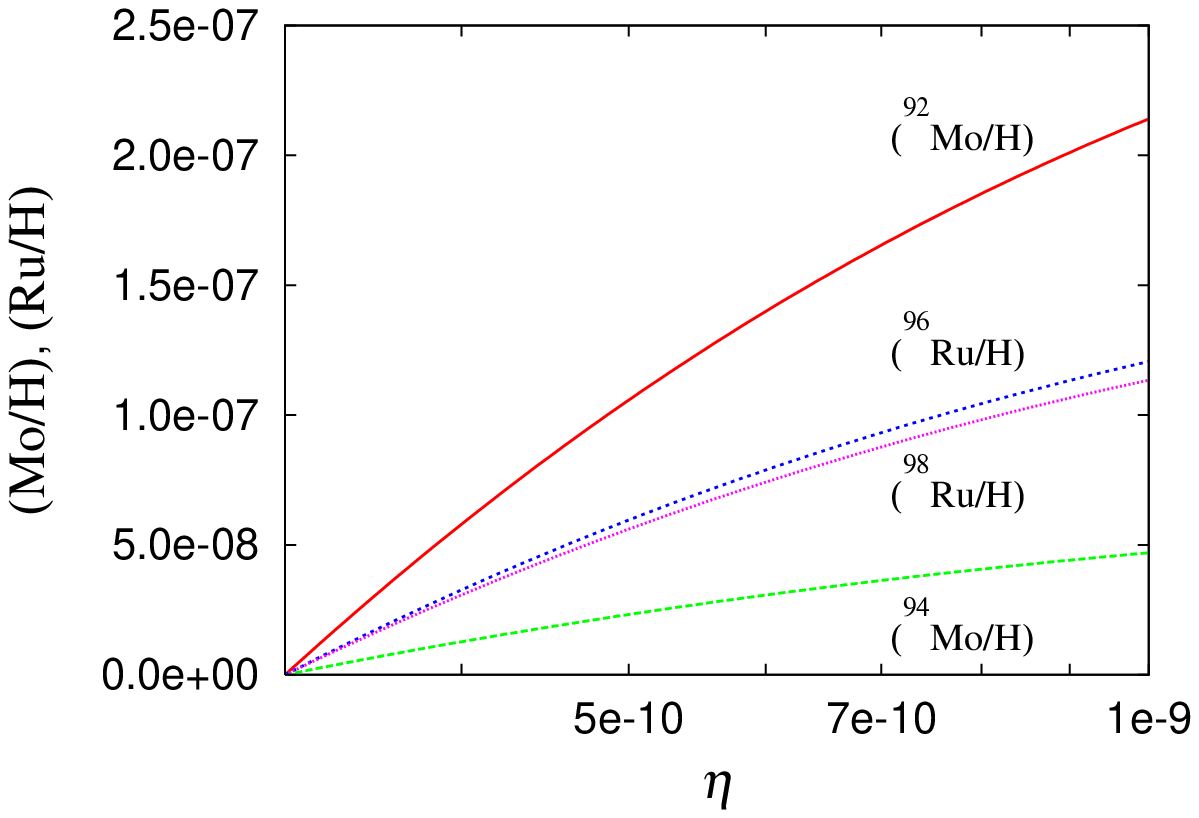}
\caption{($^{92}$\rm{Mo/H}),($^{94}$\rm{Mo/H}),($^{96}$\rm{Ru/H}) and ($^{98}$\rm{Ru/H}) vs $\eta$. Red, green, blue and pink lines represent the ratio ($^{92}$\rm{Mo/H}),($^{94}$\rm{Mo/H}),($^{96}$\rm{Ru/H}),($^{98}$\rm{Ru/H}) respectively.}
\label{combps}
\end{figure}

We plot expected value of these quantities in Fig.\ref{combps}.

%
%
%
%

These values should be compared with the solar abundance(Table.\ref{solar})\cite{Anders:1989zg}.
\begin{center}
\begin{table}[htbp]
\begin{tabular}{ccc}\hline
name & number fraction & ratio to H \\ \hline
H & $7.057280 \times 10^{-1}$ & 1 \\
$^{92}$Mo & $8.796560 \times 10^{-10}$ & $1.2465 \times 10^{-9} $ \\
$^{94}$Mo &$ 5.611420 \times 10^{-10}$ & $ 7.9512\times 10^{-10} $  \\
$^{96}$Ru & $2.501160 \times 10^{-10}$ & $ 3.5441\times 10^{-10} $ \\
$^{98}$Ru &$ 8.676150 \times 10^{-11} $&  $ 1.2294\times 10^{-10} $ \\ \hline
\end{tabular}
\caption{The abundances of $^{92,94}$\rm{Mo} and $^{96,98}$\rm{Ru} in the solar system\cite{Anders:1989zg}}
\label{solar}
\end{table}
\end{center}

Compared those observational values with Fig.\ref{combps},
it is clear that the heavy element produced in BBN can affect the solar abundance heavy element.
Some of them are produced too much. But this is not a problem of the previous work
\cite{Matsuura:2005rb}, because we assumed that high density regions are very small and 
do not disturb standard BBN.
The analysis here suggest that even if we assume the density fluctuations are completely mixed,
heavy element can have enough affect to the solar abundance.

\section{Diffusion during BBN}
\label{DIFF}

In the previous analysis, we assumed that the diffusion effect can be ignored during BBN and 
both high density regions and low density regions are completely mixed after BBN. 
In this section, we determine the scale of high baryon density island in which the 
diffusion effect during BBN is very small enough and our assumption is valid.
We do not discuss the diffusion after BBN here.

A detail analysis of the comoving diffusion distance of the baryon, the neutron and the proton is in \cite{Applegate:1987hm}.
From Fig.1 in \cite{Applegate:1987hm}, in order to safely ignore the diffusion effect, it is necessary for 
the high baryon density island to be much larger than $10^{5}$cm at T=0.1MeV($1.1\times 10^{9}$K).
Notice that $T \propto \frac{1}{A}$, where A is a scale factor.
For scale d now corresponds to $d/(4.0\times10^8)$ at BBN epoch.
Present galaxy scale is $\mathcal{O}(10^{20})$cm, which corresponds to
$\mathcal{O}(10^{12})$cm $>>10^{5}$cm at BBN epoch.

\begin{center}
\begin{table}[htbp]
\begin{tabular}{cc}\hline
\multicolumn{2}{c}{temperature and scale} \\ \hline
temperature  & scale \\
$1.1\times10^{9}$K (BBN) & d \\
3000K (decouple) & $3.7\times 10^{6}\times d$ \\
2.725K (now) & $4.0\times 10^{8}\times d$ \\ \hline
\end{tabular}
\caption{Relation between temperature and scale}
\end{table}
\end{center}

The maximum angular resolution of CMB is $l_{max} \sim$2000.
The size of universe is $\sim 5000$Mpc.
In order not to contradict to CMB observation, the fluctuation of baryon density must be
less than $\sim 16$Mpc now.
This corresponds to $10^{17}$cm at BBN.

Since the density fluctuation size in Dolgov and Silk's model\cite{Dolgov:1992pu} is a free parameter,
the above brief estimation suggests that we can take the island size large enough to ignore the diffusion effect
without contradicting to observations, i.e., the reasonable size of $10^{5}$cm $-$$10^{17}$cm at the BBN epoch.
We can choose distances between high density islands so that we obtain a suitable value of $f$.


\section{Summary}
In this paper, we studied the relation between the heavy element production in high baryon density regions during BBN
and the light element observation.
By averaging the light element abundances in the high and the low density regions we showed that  it is possible to produce a relevant amount of heavy element without contradicting to observations.
However we should stress that in this paper we restricted ourselves 
to some parameter regions where neutrons in high baryon density regions do not destroy the standard BBN. 
So our setup is different from the conventional inhomogeneous
BBN studies.
We also studied the size of the density fluctuation to show that there is a parameter region in which the neutron diffusion is negligible
and which is much smaller than CMB observation scale.
It is worthwhile to investigate further how the produced heavy elements
can be related to the detailed observations.

\section{Acknowledgements}
We thank R.H. Cyburt, R. Allahverdi and R. Nakamura for useful discussions.
This research was supported in part by Grants-in-Aid for Scientific
Research provided by the Ministry of Education, Science and 
Culture of Japan through Research Grant No.S 14102004, No.14079202. 
S.M.'s work was supported in part by JSPS(Japan Society for the
Promotion of Science).

\end{document}